\documentclass[a4paper, 10pt, twosides]{amsart}

\usepackage{lmodern}
\usepackage{amssymb}
\usepackage{amsthm}
\usepackage{amsaddr}
\usepackage{mathtools}
\usepackage{MnSymbol}
\usepackage{graphicx}
\usepackage{caption}
\usepackage{subcaption}
\usepackage{url}

\usepackage{enumitem}
\setitemize{nolistsep}
\setenumerate{nolistsep}
\usepackage{todonotes}
\usepackage{algorithm,algorithmic}

\allowdisplaybreaks

\newtheorem{theorem}{Theorem}

\theoremstyle{definition}
\newtheorem{defn}{Definition}

\theoremstyle{remark}
\newtheorem{remark}{Remark}

\theoremstyle{assumption}

\theoremstyle{fact}
\newtheorem{fact}{Fact}

\theoremstyle{claim}

\theoremstyle{prob}
\newtheorem{prob}{Problem}

\theoremstyle{algo}

\theoremstyle{experiment}
\newtheorem{experiment}{Experiment}

\numberwithin{equation}{section}
\newcommand{\abs}[1]{\left\lvert{#1}\right\rvert}

\newcommand{\norm}[1]{\left\lVert#1\right\rVert}
\newcommand{\pmat}[1]{\begin{pmatrix}#1\end{pmatrix}}

\newcommand{\R}{\mathbb{R}}
\newcommand{\N}{\mathbb{N}}

\renewcommand{\P}{\mathcal{P}}

\newcommand{\M}{\mathcal{M}}

\title[]{On the design of stabilizing cycles for\\switched linear systems}
\author{Atreyee Kundu}
\address{Department of Electrical Engineering,\\Indian Institute of Science Bangalore,\\Bengaluru - 560012, India,\\ E-mail: atreyeek@iisc.ac.in}


\date{\today}

\begin{document}

	\begin{abstract}
       Given a family of systems, identifying stabilizing switching signals in terms of infinite walks constructed by concatenating cycles on the underlying directed graph of a switched system that satisfy certain conditions, is a well-known technique in the literature. This paper deals with a new {method to design} these cycles for stability of switched linear systems. We employ properties of the subsystem matrices and mild assumption on the admissible switches between the subsystems {for this purpose}. In contrast to prior works, {our construction of} stabilizing cycles does not involve design of Lyapunov-like functions and storage of sets of scalars in memory prior to the application of a cycle detection algorithm. As a result, {the} techniques {proposed in this paper} offer improved numerical tractability.
    \end{abstract}

    \maketitle
\section{Introduction}
\label{s:intro}
   \paragraph{Motivation}
    A \emph{switched system} has two ingredients --- a family of systems and a switching signal. The \emph{switching signal} selects an \emph{active subsystem} at every instant of time, i.e., the system from the family that is currently being followed \cite[\S 1.1.2]{Liberzon2003}. In this paper we work with switched systems in the setting of linear dynamics in discrete-time.

    Given a family of systems, {identification of} classes of switching signals that preserve stability of the resulting switched system is a key topic in the literature \cite[Chapter 3]{Liberzon2003}. In the recent past this problem has been studied widely by employing multiple Lyapunov-like functions  \cite{Branicky1998} and graph-theoretic tools. A weighted directed graph \cite{Bollobas} is associated to a family of systems and the admissible transitions between them; the vertices of this graph are weighted by a measure of the rate of growth or decay of the Lyapunov-like functions corresponding to the subsystems, and the edges are weighted by a measure of the jump between these functions. A switching signal is expressed as an infinite walk on the above directed graph. Infinite walks whose corresponding switching signals preserve stability, are {constructed} as concatenation of negative weight cycles \cite{Bollobas} (henceforth, also referred to as stabilizing cycles).\footnote{By the term `stabilizing cycles' we refer to cycles such that a switching signal corresponding to an infinite walk constructed by concatenating these cycles, is stabilizing. We will discuss and employ different {methods of constructing} such cycles throughout the paper.} This class of results was first introduced in \cite{KunCha_HSCC_2014} for switched linear systems, and was later extended to the setting of switched nonlinear systems in \cite{KunCha_NAHS_2017}. A primary feature of the stability conditions proposed in \cite{KunCha_HSCC_2014, KunCha_NAHS_2017} is their numerical tractability compared to the prior results that rely on point-wise properties of the switching signals \cite{Zhai2002,KunMisCha_CDC_2015}. In contrast to verifying certain conditions on the number of switches and duration of activation of unstable subsystems on \emph{every} interval of time, one needs to detect negative weight cycles on the underlying weighted directed graph of a switched system.

    Existence of these cycles depends on two factors: (i) connectivity of the directed graph, and (ii) weights associated to the vertices and edges of this graph. While (i) is determined by the admissible switches between the subsystems, the elements of (ii) are scalar quantities calculated from Lyapunov-like functions, the choice of which is not unique. This feature leads to the problem of `co'-designing Lyapunov-like functions such that the underlying weighted directed graph of a switched system admits a stabilizing cycle. This design problem, in general, is numerically difficult, see \cite[\S 3]{KunCha_NAHS_2017} for a detailed discussion.

    As a natural choice, the existing literature considers the Lyapunov-like functions and the corresponding vertex and edge weights to be ``given'', and detects negative weight cycles on the underlying weighted directed graph of a switched system. However, non-existence of a negative weight cycle with the given choice of vertex and edge weights does not conclude that a family of systems does not admit such cycles at all. In addition, storing the vertex and edge weights prior to the application of a negative weight cycle detection algorithm requires a huge memory when the number of subsystems is large. These features motivate the search for a {method to construct} stabilizing cycles {that is} independent of the choice of Lyapunov-like functions and {works} without complete knowledge of the vertex and edge weights of the underlying directed graph of a switched system. In this paper we report our result that addresses these requirements.

    \paragraph{Our contributions}
    Given a family of systems (containing both stable and unstable components) and a set of admissible switches between the subsystems, we {first identify sufficient conditions on} cycles on the underlying directed graph of the switched system {such that they are stabilizing}. {These conditions are derived} in terms of properties of the subsystem matrices, and hence the use of multiple Lyapunov-like functions is avoided. In particular, we rely on (matrix) commutators between the subsystem matrices and mild assumption on the admissible switches between the subsystems for this purpose. Our stability condition involves the rate of decays of the Schur stable matrices, upper bounds on the Euclidean norms of the commutators of the subsystem matrices, certain scalars capturing the properties of these matrices individually, and the total number of subsystems. In contrast to prior works, our directed graphs are unweighted, and the detection of our stabilizing cycles depends only on the activation of a vertex whose corresponding subsystem satisfies certain conditions. We {then} present an algorithm to detect cycles {on the underlying directed graph of a switched system} that satisfy our conditions.

    Matrix commutators (Lie brackets) have been employed to study stability of switched systems with all or some stable subsystems earlier in the literature \cite{Narendra1994, Agrachev2012, abc, def}. However, to the best of our knowledge, this is the first instance when they are employed to {construct} stabilizing cycles for switched systems.

    In summary, the main contribution of this paper is in extending the technique of employing stabilizing cycles to {construct} stabilizing switching signals, by proposing a new {method for designing} these cycles. {Our} class of stabilizing cycles offers better numerical tractability in terms of detecting its elements compared to the existing results.

    \paragraph{Paper organization}
    The remainder of this paper is organized as follows: we formulate the problem under consideration and catalog the required preliminaries for our result in \S\ref{s:prelims}. Our main result appears in \S\ref{s:mainres}. We also discuss various features of our result in this section. In \S\ref{s:num_ex} we present numerical experiments. We provide a comparison of our result with the existing methods in \S\ref{s:discssn}, and conclude in \S\ref{s:concln}. A proof of our main result appears in \S\ref{s:all_proofs}.

    \paragraph{Notation}
    \(\N\) is the set of natural numbers, \(\N_{0} = \N\cup\{0\}\). \(\norm{\cdot}\) denotes the Euclidean norm (resp., induced matrix norm) of a vector (resp., a matrix). For a matrix \(P\), given by a product of matrices \(M_{i}\)'s, \(\abs{P}\) denotes the length of the product, i.e., the number of matrices that appear in \(P\), counting repetitions.
\section{Preliminaries}
\label{s:prelims}
\setcounter{paragraph}{0}
   \paragraph{The problem}
   We consider a discrete-time switched linear system \cite[\S 1.1.2]{Liberzon2003}
    \begin{align}
    \label{e:swsys}
        x(t+1) = A_{\sigma(t)}x(t),\:\:x(0)=x_{0},\:\:t\in\N_{0}
    \end{align}
    generated by
    \begin{itemize}[label = \(\circ\), leftmargin = *]
        \item a family of systems
        \begin{align}
        \label{e:family}
            x(t+1) = A_{i}x(t),\:\:x(0) = x_{0},\:\:i\in\P,\:\:t\in\N_{0},
        \end{align}
        where \(x(t)\in\R^{d}\) is the vector of states at time \(t\), \(\P = \{1,2,\ldots,N\}\) is an index set, \(A_{i}\in\R^{d\times d}\), \(i\in\P\) are constant matrices, and
        \item a switching signal \(\sigma:\N_{0}\to\P\) that specifies at every time \(t\), the index of the {active subsystem}, i.e., the dynamics from \eqref{e:family} that is being followed at \(t\).
    \end{itemize}

    The solution to \eqref{e:swsys} is given by
    \begin{align}
    \label{e:soln}
        x(t) = A_{\sigma(t-1)}\ldots A_{\sigma(1)}A_{\sigma(0)}x_{0},\:\:t\in\N,
    \end{align}
    where we have suppressed the dependence of \(x\) on \(\sigma\) for notational simplicity.

    \begin{defn}{\cite[\S 2]{Agrachev2012}}
	 \label{d:ges}
    \rm{
	 	The switched system \eqref{e:swsys} is \emph{globally exponentially stable (GES) under a switching signal \(\sigma\)} if there exist positive numbers \(c\) and \(\gamma\) such that for arbitrary choices of the initial condition \(x_{0}\), the following inequality holds:
		\begin{align}
		\label{e:ges1}
            \norm{x(t)}\leq c\exp(-\gamma t)\norm{x_{0}}\:\:\text{for all}\:\:t\in\N.
		\end{align}
        }
	 \end{defn}

    We are interested in the following problem:
    \begin{prob}
    \label{prob:mainprob}
        Given a family of systems \eqref{e:family} (containing both stable and unstable components) and a set of admissible switches, find a class of switching signals \(\mathcal{S}\) such that the switched system \eqref{e:swsys} is GES under every \(\sigma\in\mathcal{S}\).
    \end{prob}

    Prior to presenting our solution to Problem \ref{prob:mainprob}, we catalog some required preliminaries.

   \paragraph{Family of systems}
   Let \(\P_{S}\) and \(\P_{U}\subset\P\) denote the sets of indices of Schur stable and unstable subsystems, respectively, \(\P = \P_{S}\sqcup\P_{U}\).\footnote{A matrix \(A\in\R^{d\times d}\) is Schur stable if all its eigenvalues are inside the open unit disk. We call \(A\) unstable if it is not Schur stable.} Let \(E(\P)\) denote the set of ordered pairs \((i,j)\) such that a switch from subsystem \(i\) to subsystem \(j\) is admissible, \(i,j\in\P\).\footnote{Clearly, \((i,i)\) implies that it is allowed to dwell on a subsystem for two or more consecutive time steps.}

    Let
    \begin{align}
    \label{e:M_defn}
        M = \max_{i\in\P}\norm{A_{i}}.
    \end{align}

    The following fact follows from the properties of Schur stable matrices:
    \begin{fact}
    \label{fact:key}
        There exists \(m\in\N\) and \(\rho\in]0,1[\) such that the following set of inequalities holds:
        \begin{align}
        \label{e:m_ineq}
            \norm{A_{i}^{m}}\leq\rho,\:\:i\in{\P_{S}}.
        \end{align}
    \end{fact}

    We will employ the set of (matrix) commutators defined below in our stability analysis:\footnote{These matrix commutators will be employed in rearranging the matrix products on the right-hand side of \eqref{e:soln}. An exchange between \(A_{p}\) with itself is undefined, and hence \(i\in\P\setminus\{p\}\).}
    \begin{align}
    \label{e:comm}
        E_{p,i} = A_{p}A_{i} - A_{i}A_{p},\:\:p\in\P_{S},\:\:i\in \P\setminus\{p\}.
    \end{align}

    \paragraph{Switching signals}
    We associate a directed graph \(G(V,E)\) with the switched system \eqref{e:swsys} in the following manner:
    \begin{itemize}[label = \(\circ\), leftmargin = *]
        \item The set of vertices \(V\) is the set of indices of the subsystems, \(\P\).
        \item The set of edges \(E\) contains a directed edge from a vertex \(i\) to a vertex \(j\) whenever \((i,j)\in E(\P)\).
    \end{itemize}
    Recall that a \emph{walk} on a directed graph is an alternating sequence of vertices and edges \(W = v_{0}, e_{1}, v_{1}, e_{2}, v_{2}, \ldots, v_{n-1}, e_{n}, v_{n}\), where \(v_{\ell}\in V\), \(e_{\ell} = (v_{\ell-1},v_{\ell})\in E\), \(0 < \ell \leq n\). The length of a walk is its number of edges, counting repetitions, e.g., the length of \(W\) above is \(n\). The \emph{initial vertex} of \(W\) is \(v_{0}\) and the \emph{final vertex} of \(W\) is \(v_{n}\). If \(v_{0}=v_{n}\), we say that the walk \(W\) is closed. A closed walk \(W\) is called a \emph{cycle} if the vertices \(v_{i}\), \(0<i<n\) are distinct from each other and \(v_{0}\). By the term \emph{infinite walk} we mean a walk of infinite length, i.e., it has infinitely many edges.
    \begin{fact}{\cite[Fact 3]{KunCha_HSCC_2014}}
    \label{fact:walk}
        The set of switching signals \(\sigma:\N_{0}\to\P\) and the set of infinite walks on \(G(V,E)\) are in bijective correspondence.
    \end{fact}
    Clearly, given a family of systems \eqref{e:family} and a set of admissible switches, we are interested in a class of infinite walks whose corresponding class of switching signals ensures GES of the switched system \eqref{e:swsys}.

    For a cycle \(W\) on \(G(V,E)\), let
    \(
        \mathcal{V}(W) = \{v\in V\:|\:v\:\:\text{appears in}\:\:W\},\:\text{and}
    \)
    \(
        \mathcal{E}(W) = \{(u,v)\in E\:|\:(u,v)\:\:\text{appears in}\:\:W\}.
    \)

    \begin{defn}
    \label{d:v-cycles}
        Fix a vertex \(\overline{v}\in V\). A cycle \(W = v_{0},(v_{0},v_{1}),v_{1},\ldots,v_{n-1},(v_{n-1},v_{0}),v_{0}\) on \(G(V,E)\) is called a \(\overline{v}\)-cycle if \(\overline{v}\in\mathcal{V}(W)\).
    \end{defn}

    \begin{defn}
    \label{d:concat_cycle}
         The distinct cycles
         \(
            W_{1} = v_{0}^{(1)}, (v_{0}^{(1)},v_{1}^{(1)}), v_{1}^{(1)}\),\\\(\ldots,v_{n_{1}-1}^{(1)}, (v_{n_{1}-1}^{(1)},v_{0}^{(1)}), v_{0}^{(1)},
    W_{2} = v_{0}^{(2)}, (v_{0}^{(2)},v_{1}^{(2)}), v_{1}^{(2)},\ldots\),\\\(v_{n_{1}-1}^{(2)}, (v_{n_{1}-1}^{(2)},v_{0}^{(2)}), v_{0}^{(2)},\ldots,
        W_{r} = v_{0}^{(r)}, (v_{0}^{(r)},v_{1}^{(r)}), v_{1}^{(r)},\ldots\),\\\(v_{n_{1}-1}^{(r)}, (v_{n_{1}-1}^{(r)},v_{0}^{(r)}), v_{0}^{(r)},
        \)
        on \(G(V,E)\) are called \emph{concatenable} on a vertex \(u\in V\) if \(v_{0}^{(1)} = v_{0}^{(2)}\) = \(\cdots\) = \(v_{0}^{(r)}\) = \(u\).
    \end{defn}

    Let \(C_{\overline{v}}\) denote the set of all \(\overline{v}\)-cycles on \(G(V,E)\), \(\overline{v}\in V\), that are concatenable on vertex \(u\) for some \(u\in V\). We will consider the following structures:
        \begin{itemize}[label = \(\circ\), leftmargin = *]
            \item If there is no cycle \(W\) on \(G(V,E)\) such that \(\overline{v}\in\mathcal{V}(W)\), then \(C_{\overline{v}} = \emptyset\).
            \item If \(G(V,E)\) admits exactly one \(\overline{v}\)-cycle \(W\), then \(C_{\overline{v}}\) is a singleton. In particular, \(C_{\overline{v}} = \{W\}\).
        \end{itemize}

    We will work with infinite walks on \(G(V,E)\) constructed by concatenating the elements from \(C_{\overline{v}}\). At this point, it is important to clarify the meaning of such a concatenation. Let \(W_{1}, W_{2}, \ldots, W_{q}\) be distinct concatenable \(\overline{v}\)-cycles of length \(n_{1},n_{2},\ldots,n_{q}\) on \(G(V,E)\), respectively. An infinite walk \(W\) obtained by concatenating \(W_{j}\), \(j = 1,2,\ldots,q\) is
       \( W = v_{0}^{(j_{1})}, \bigl(v_{0}^{(j_{1})}, v_{1}^{(j_{1})}\bigr), v_{1}^{(j_{1})}, \ldots, v_{n_{1}-1}^{(j_{1})}, \bigl(v_{n_{1}-1}^{(j_{1})},v_{0}^{(j_{1})}\bigr),
         v_{0}^{(j_{2})}, \bigl(v_{0}^{(j_{2})}\),\\\(v_{1}^{(j_{2})}\bigr), v_{1}^{(j_{2})}, \ldots, v_{n_{2}-1}^{(j_{2})}, \bigl(v_{n_{2}-1}^{(j_{2})},v_{0}^{(j_{2})}\bigr),
         v_{0}^{(j_{3})},\ldots\),\\\(j_{k}\in\{1,2,\ldots,q\},\:\:k\in\N.\)

    We are now in a position to present our result.
\section{Result}
\label{s:mainres}
    \begin{theorem}
    \label{t:mainres}
    {\it
        Consider a switched system \eqref{e:swsys} and its underlying directed graph \(G(V,E)\). Let \(\gamma\) be an arbitrary positive number such that
    \begin{align}
    \label{e:maincondn1}
        \rho e^{\gamma m} < 1.
    \end{align}
        Suppose that there exists \(p\in\P_{S}\) that satisfies the following conditions:
        \begin{itemize}[label = \(\circ\), leftmargin = *]
            \item \(C_{p} \neq \emptyset\), and
            \item there exists a scalar \(\varepsilon_{p}\) small enough such that
            \begin{align}
            \label{e:maincondn2}
                \norm{E_{p,i}}\leq\varepsilon_{p}\:\:\text{for all}\:i\in\P\setminus\{p\},
            \end{align}
            and
            \begin{align}
            \label{e:maincondn3}
                \rho e^{\gamma m} + (N-1)\frac{m(m+1)}{2}M^{Nm-2}\varepsilon_{p}e^{\gamma Nm}\leq 1.
            \end{align}
        \end{itemize}
        Then the switched system \eqref{e:swsys} is GES under every switching signal \(\sigma\) whose corresponding infinite walk \(W\) is constructed by concatenating elements from \(C_{p}\).
    }
    \end{theorem}

    Theorem \ref{t:mainres} is our solution to Problem \ref{prob:mainprob}. The elements of the class of stabilizing switching signals \(\mathcal{S}\) correspond to the infinite walks that are constructed by concatenating elements from the set of cycles, \(C_{p}\), where \(p\) is a stable subsystem satisfying certain conditions. Notice that since \(\rho < 1\), there always exists a number \(\gamma\) (could be very small) such that condition \eqref{e:maincondn1} holds. If in addition, there exists a \(p\in\P_{S}\) such that \(G(V,E)\) admits at least one cycle that involves vertex \(p\), and the Euclidean norms of (matrix) commutators of \(A_{p}\) and \(A_{i}\), \(i\in\P\setminus\{p\}\), are bounded above by a scalar \(\varepsilon_{p}\) small enough such that condition \eqref{e:maincondn3} holds, then \eqref{e:swsys} is GES under a \(\sigma\) whose corresponding infinite walk is constructed by concatenating the cycles from \(C_{p}\). Our stability conditions have the following important features:\\
        1) Theorem \ref{t:mainres} accommodates sets of matrices \(A_{j}\), \(j\in\P\), for which \(A_{p}\) and \(A_{i}\), \(p\in\P_{S}\), \(i\in\P\setminus\{p\}\), do not necessarily commute, but are ``close'' to sets of matrices for which they commute. When these matrices commute for all \(i\in\P\setminus\{p\}\), (i.e., \(\varepsilon_{p} = 0\)), condition \eqref{e:maincondn3} reduces to condition \eqref{e:maincondn1}. The stability conditions of Theorem \ref{t:mainres} are inherently robust in the above sense. Indeed, if we are relying on approximate models of \(A_{j}\), \(j\in\P\), or the elements of \(A_{j}\), \(j\in\P\), are prone to evolve over time, then GES of \eqref{e:swsys} holds under our stabilizing switching signals as long as for some \(p\in\P_{S}\), the commutators of \(A_{p}\) and \(A_{i}\), \(i\in\P\setminus\{p\}\), in their Euclidean norm, are bounded above by a small scalar \(\varepsilon_{p}\) such that condition \eqref{e:maincondn3} holds.\\
        2) {The proposed construction of infinite walks} involves activation of a vertex (subsystem) \(p\in\P_{S}\) for which conditions \eqref{e:maincondn2}-\eqref{e:maincondn3} hold. Clearly, we require \(\P_{S}\neq\emptyset\). However, stabilizing cycles with all unstable vertices except for \(p\) are perfectly admissible. The activation of at least one Schur stable subsystem is also a requirement for Lyapunov-like functions based {construction} of stabilizing cycles.

    \begin{remark}
    \label{rem:comm_compa}
        Commutation relations between the subsystem matrices or certain products of these matrices have been employed to study stability of the switched system \eqref{e:swsys} earlier in the literature. The switched system \eqref{e:swsys} is stable under arbitrary switching \cite[Chapter 2]{Liberzon2003} if all the subsystems \(A_{i}\), \(i\in\P\), are Schur stable, and commute pairwise \cite{Narendra1994} or are ``sufficiently close'' to a set of matrices whose elements commute pairwise \cite{Agrachev2012}. Recently in \cite{abc,def} stability of \eqref{e:swsys} under restricted switching is studied using matrix commutators. Given an admissible minimum dwell time, in \cite{abc} we identify conditions on the subsystem matrices such that stability of \eqref{e:swsys} is preserved under all switching signals satisfying the given minimum dwell time. The overarching hypothesis there is that all subsystems are Schur stable. The problem of identifying classes of stabilizing switching signals when not all systems in the family \eqref{e:family} are stable and an admissible switching signal obeys certain minimum and maximum dwell times on all subsystems, is addressed in \cite{def}. In the current work we tackle the problem of {algorithmically constructing} stabilizing cycles by employing commutation relations between the subsystem matrices. We deal with families containing both stable and unstable subsystems, and obeying pre-specified restrictions on admissible switches between the subsystems. See Remark \ref{rem:proof_tech} for a discussion on our analysis technique.
    \end{remark}

    We next provide an algorithm (Algorithm \ref{algo:sw-sig_construc}) that detects a vertex (subsystem) \(p\) such that conditions \eqref{e:maincondn2}-\eqref{e:maincondn3} hold, and {constructs} the set of cycles, \(C_{p}\).
Once a suitable \(C_{p}\) is obtained, an infinite walk \(W = v_{0}, (v_{0},v_{1}),v_{1}, (v_{1},v_{2}),v_{2},\ldots\) can be constructed by concatenating the elements of \(C_{p}\) as described in \S\ref{s:prelims}, and the corresponding switching signal \(\sigma\) can be designed as \(\sigma(0) = v_{0}\), \(\sigma(1) = v_{1}\), \(\sigma(2) = v_{2},\:\:\ldots\).
        Clearly, if \(C_{p}\) is a singleton, e.g., \(C_{p} = \{W\}\), then a stabilizing \(\sigma\) described in Theorem \ref{t:mainres} is periodic. It corresponds to the infinite walk constructed by repeating \(W\).

        \begin{algorithm}[htbp]
			\caption{Detection of \(p\) for which conditions \eqref{e:maincondn2}-\eqref{e:maincondn3} hold, and construction of \(C_{p}\)}\label{algo:sw-sig_construc}
		\begin{algorithmic}[1]
			\renewcommand{\algorithmicrequire}{\textbf{Input:}}
			\renewcommand{\algorithmicensure}{\textbf{Output:}}
			
			\REQUIRE a family of systems \eqref{e:family} and a set of admissible switches \(E(\P)\)		
            \ENSURE a set of concatenable \(p\)-cycles, \(C_{p}\)
			
			 \STATE Construct the underlying directed graph \(G(V,E)\) of the switched system \eqref{e:swsys}.
			  \STATE Compute \(m\) and \(\rho\) such that \eqref{e:m_ineq} holds.
                \STATE Compute \(\gamma\) such that \eqref{e:maincondn1} holds.

             \FOR {\(p\in\P_{S}\)}
                \STATE Compute \(\varepsilon_{p}\) such that \eqref{e:maincondn2} holds.
                \IF {\eqref{e:maincondn3} is satisfied,}
                    \STATE Create a set \(L_{p}\) containing all \(p\)-cycles on \(G(V,E)\).
                    \IF {\(L_{p}\neq\emptyset\),}
                        \IF {\(L_{p}\) contains concatenable elements,}
                            \STATE Create a set \(C_{p}\) with all elements of \(L_{p}\) that are concatenable on a vertex \(u\) for some \(u\in V\).
                        \ELSE
                            \STATE Create a set \(C_{p}\) with any one element of \(L_{p}\).
                        \ENDIF
                    \ELSE
                        \STATE \(C_{p} = \emptyset\).
                    \ENDIF
                    \IF {\(C_{p}\neq\emptyset\)}
                        \STATE Store \(C_{p}\) in memory.
                        \STATE Exit Algorithm \ref{algo:sw-sig_construc}
                    \ENDIF
                \ENDIF
			 \ENDFOR
		\end{algorithmic}
	\end{algorithm}

            An important aspect of our stability conditions is the existence and detection of a set of cycles, \(C_{p}\), corresponding to a vertex (subsystem) \(p\) for which conditions \eqref{e:maincondn2}-\eqref{e:maincondn3} hold. The existence of a non-empty \(C_{p}\) depends solely on the connectivity of \(G(V,E)\) which is governed by the given set of admissible switches \(E(\P)\). We address the detection issue in two steps: first, we list out all \(p\)-cycles on \(G(V,E)\), and second, we pick elements from the above list that are concatenable on a vertex \(u\) for some \(u\in V\). Off-the-shelf algorithms from graph theory can be employed to enumerate all cycles on \(G(V,E)\) that involve a pre-specified vertex \(p\), see e.g., \cite{Johnson1975} and the references therein.

\section{Numerical experiments}
\label{s:num_ex}
\begin{experiment}
\label{ex:num_ex1}
    \rm{
    Consider \(\P = \{1,2,3\}\) with \(A_{1} = \pmat{0.86 & 0.05\\-0.07 & 0.89}\) and \(A_{2} = \pmat{0.81 & -0.07\\-0.74 & 0.73}\). Clearly, \(\P_{S} = \{1\}\) and \(\P_{U} = \{2\}\). Let \(E(\P) = \{(1,2),(2,1)\}\).

    It follows that \(L_{1} = \bigl\{\bigl(1,(1,2),2,(2,1),1\bigr),\bigl(2,(2,1),1,(1,2)\),\\\(2\bigr)\bigr\}\). Fix \(C_{1} = \{1,(1,2),2,(2,1),1\}\). It is worth noting that both \(\norm{A_{1}A_{2}}\) and \(\norm{A_{2}A_{1}}>1\), and hence stability of \eqref{e:swsys} under the switching signal \(\sigma(0) = 1\), \(\sigma(1) = 2\), \(\sigma(2) = 1\), \(\sigma(3) = 2,\ldots\) is non-trivial. We will apply the conditions of Theorem \ref{t:mainres} to determine of GES of \eqref{e:swsys} under the above \(\sigma\).

    We have \(\norm{A_{1}} = 0.89\) and \(M = 1.25\). Let \(\rho = 0.90\) and \(\gamma = 0.0001\). It follows that \(m=1\) and \(\rho e^{\gamma m} = 0.90 < 1\). We compute \(\varepsilon_{1} = \norm{A_{1}A_{2}-A_{2}A_{1}} = 0.06\). Consequently,
    \begin{align*}
		\rho e^{\gamma m} + (N-1)\frac{m(m+1)}{2}M^{Nm-2}\varepsilon_{p}e^{\gamma Nm} = 0.96 < 1.
	\end{align*}

	We demonstrate \((\norm{x(t)})_{t\in\N_{0}}\) under \(\sigma\) corresponding to 1000 different initial conditions \(x_{0}\) chosen uniformly at random from the interval \([-10,10]^{2}\) in Figure \ref{fig:x_plot}.
	\begin{figure}[htbp]
	\centering
		\includegraphics[scale=0.25]{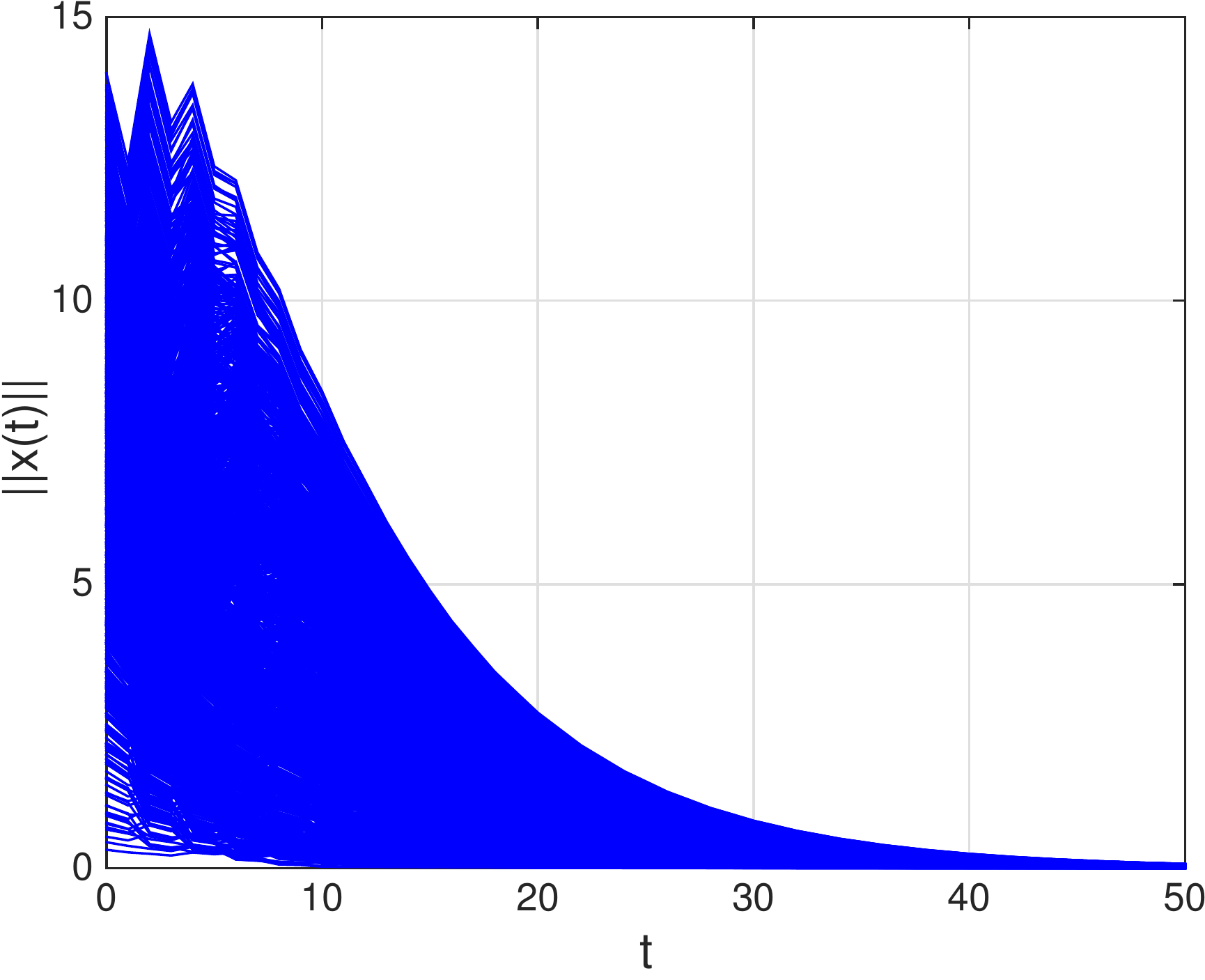}
		\caption{Plot of \(\norm{x(t)}\) versus \(t\)}\label{fig:x_plot}
	\end{figure}
    }
\end{experiment}

\begin{experiment}
\label{ex:num_ex2}
    \rm{
    We now check scalability of the stability conditions in Theorem \ref{t:mainres} with respect to \(N\), \(M\), \(m\) and \(\rho\). In the simplest case when there is a \(p\in\P_{S}\) such that the matrix \(A_{p}\) commutes with every matrix \(A_{i}\), \(i\in\P\setminus\{p\}\) (i.e., \(\varepsilon_{p} = 0\)), it follows that condition \eqref{e:maincondn3} holds even when the numbers \(M\), \(N\) and \(m\) are very large. A more interesting case is, of course, when the above matrices do not commute but are sufficiently ``close''. We fix \(\gamma = 0.0001\), and vary the parameters \(N\), \(m\), \(M\), \(\rho\) to plot upper bounds on \(\varepsilon_{p}\) for the satisfaction of condition \eqref{e:maincondn3}, see Figures \ref{fig:plot1}, \ref{fig:plot2} and \ref{fig:plot3}. Not surprisingly, it is observed that the size of the class of non-commuting matrices catered by Theorem \ref{t:mainres} shrinks with increasing \(N\), \(m\), \(M\), and \(\rho\).
    \begin{figure}[htbp]
	\centering
		\includegraphics[scale=0.25]{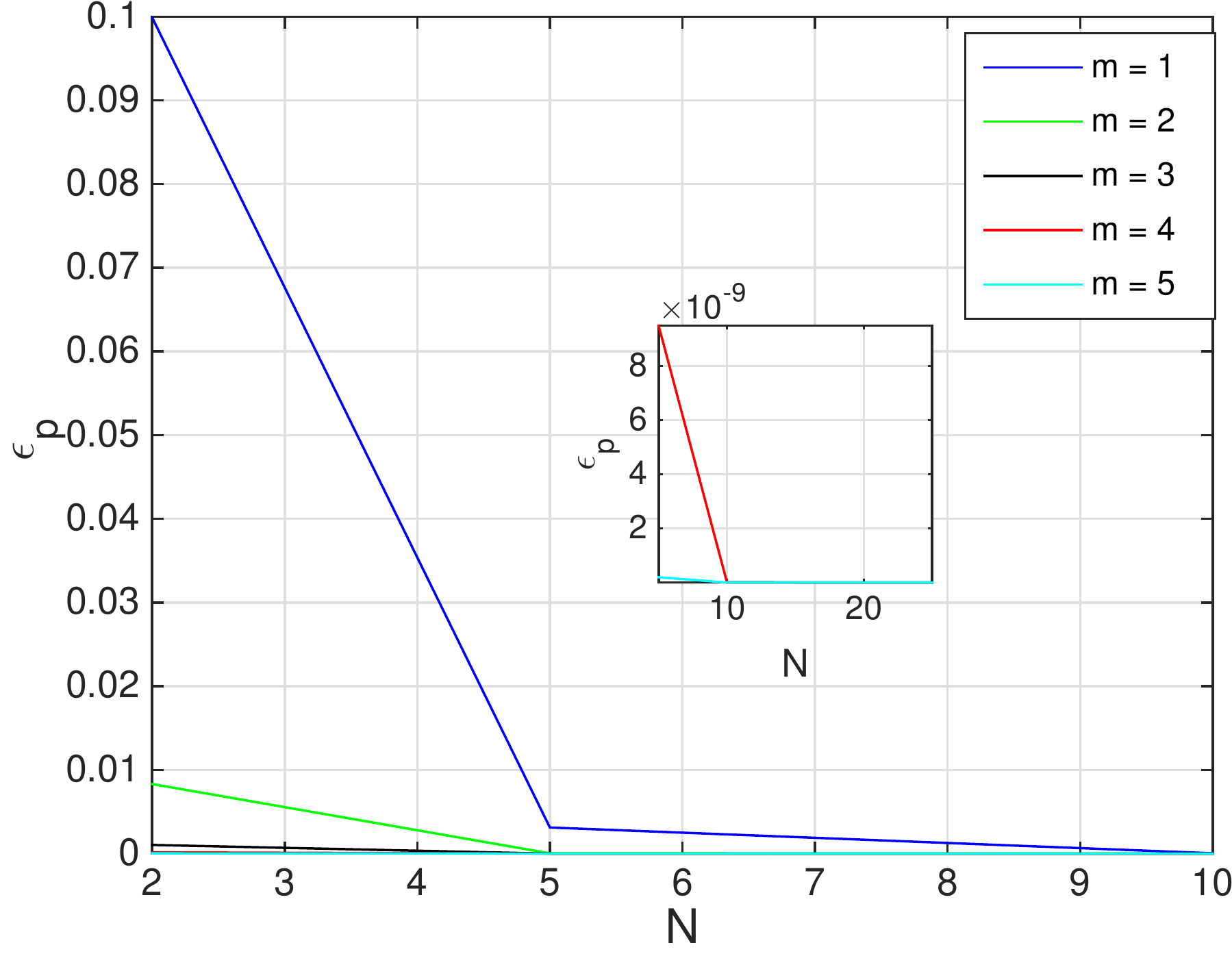}
		\caption{Plot of \(\varepsilon_{p}\) versus \(N\) with \(M = 2\) and \(\rho = 0.9\)}\label{fig:plot1}
	\end{figure}
     \begin{figure}[htbp]
	\centering
		\includegraphics[scale=0.25]{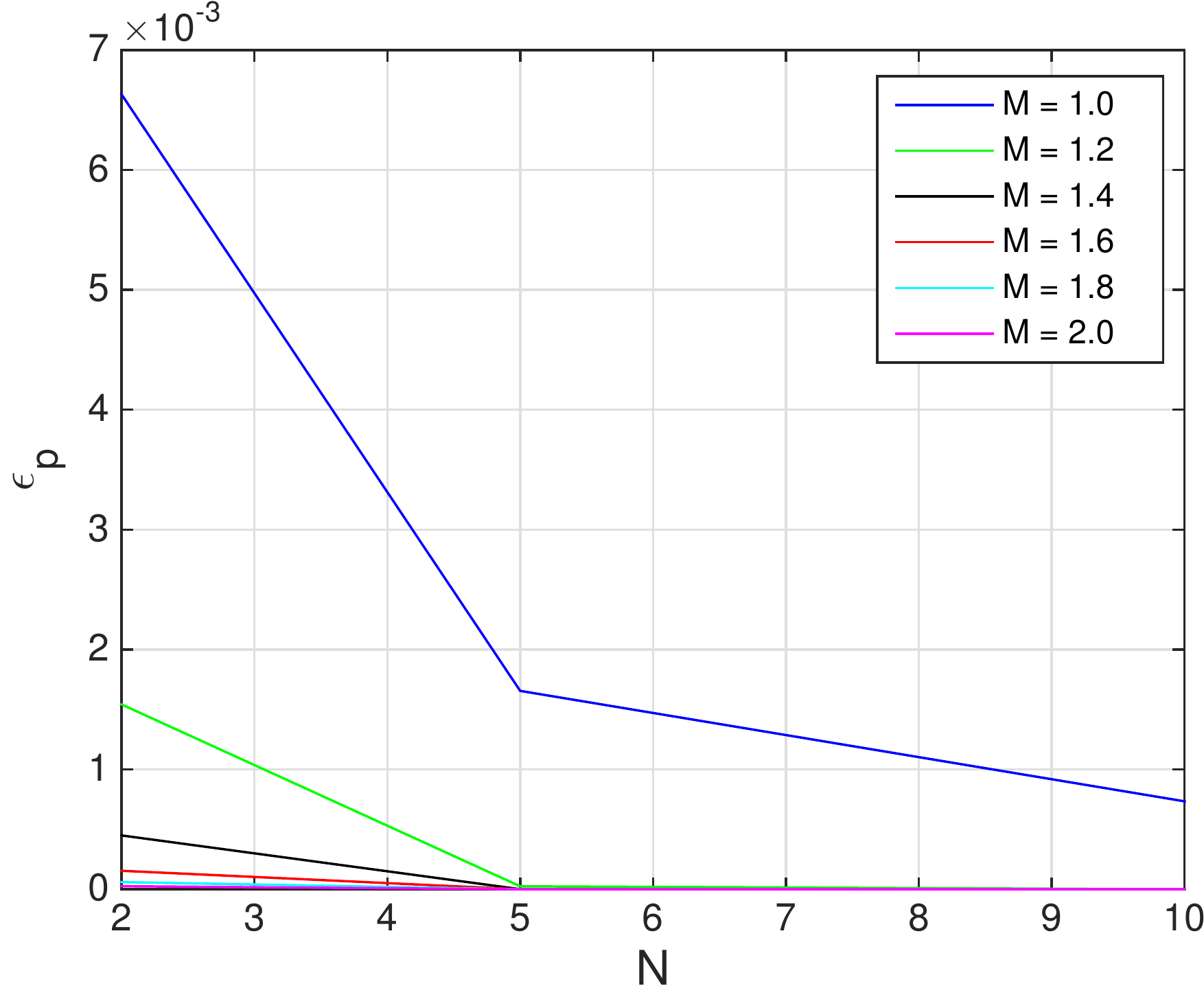}
		\caption{Plot of \(\varepsilon_{p}\) versus \(N\) with \(m = 5\) and \(\rho = 0.9\)}\label{fig:plot2}
	\end{figure}
    \begin{figure}[htbp]
	\centering
		\includegraphics[scale=0.25]{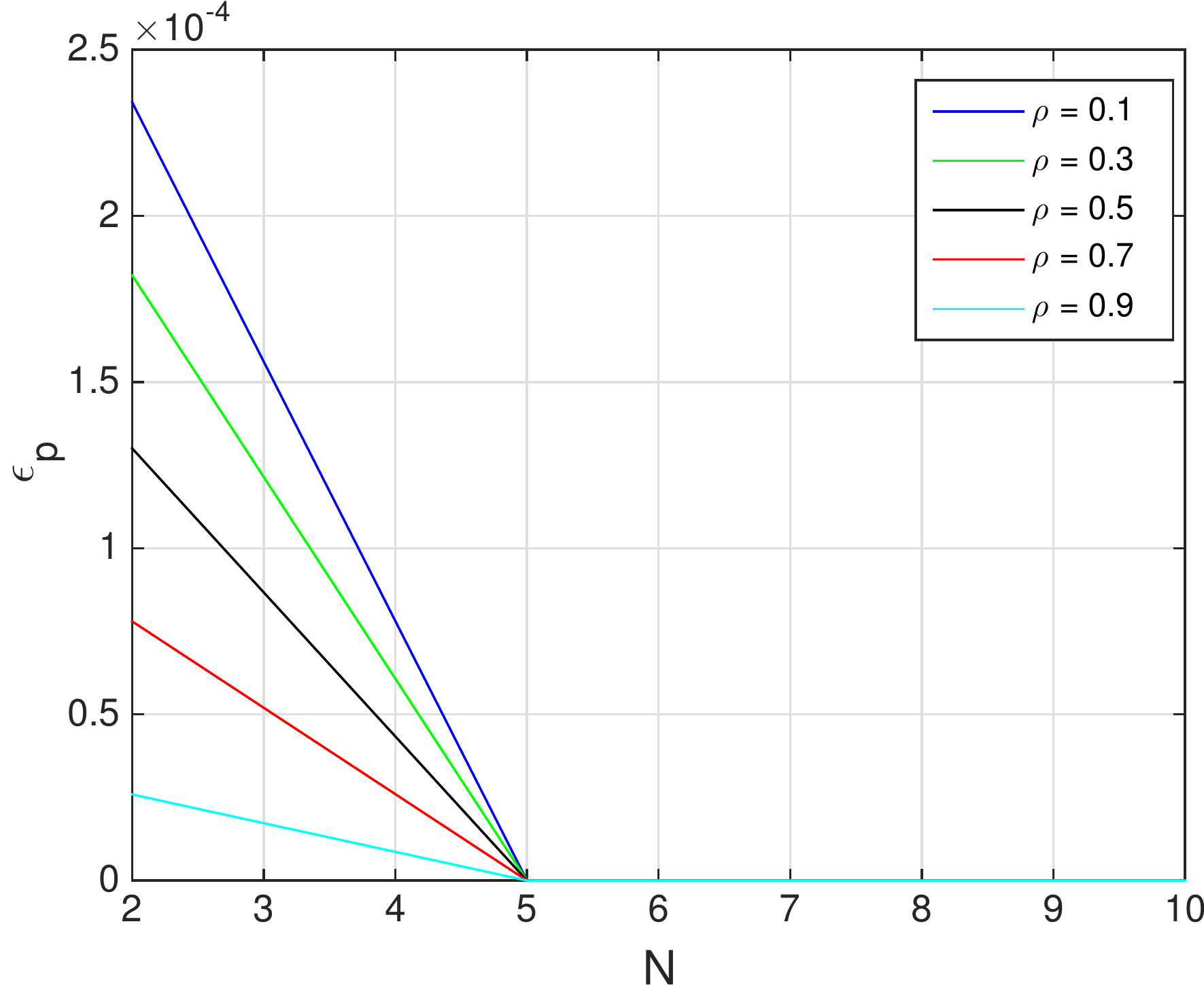}
		\caption{Plot of \(\varepsilon_{p}\) versus \(N\) with \(M = 2\) and \(m = 5\)}\label{fig:plot3}
	\end{figure}
    }
\end{experiment}
\section{Discussion}
\label{s:discssn}
        The method of characterizing stabilizing switching signals in terms of their corresponding infinite walks constructed by concatenating cycles that satisfy certain conditions, is not new in the literature. Consider a directed graph \(\overline{G}(V,E)\), which is the graph \(G(V,E)\) with vertex weights \(w(i) = -\abs{\ln\lambda_{i}},\:\text{if}\:i\in\P_{S}\) and \(w(i) = \abs{\ln\lambda_{i}},\:\text{if}\:i\in\P_{U}\), and edge weights \(w(i,j) = \ln\mu_{ij},\:(i,j)\in E(\P)\). Here the scalars \(0 < \lambda_{i} < 1\) for \(p\in\P_{S}\), \(\lambda_{i}\geq 1\) for \(p\in\P_{U}\) and \(\mu_{ij} > 0\) for \((i,j)\in E(\P)\), are computed from Lyapunov-like functions \(V_{i}:\R^{d}\to\R\) corresponding to the subsystems \(i\in\P\).

        In \cite{KunCha_HSCC_2014, KunCha_NAHS_2017} a stabilizing cycle was characterized as a negative weight cycle on \(\overline{G}(V,E)\). Notice that the existence of a negative weight cycle \(W\) on \(\overline{G}(V,E)\) depends on: (a) the connectivity of \(\overline{G}(V,E)\), and (b) the weights associated to the vertices and edges of \(\overline{G}(V,E)\). While connectivity of \(\overline{G}(V,E)\) is governed by the given set of admissible switches \(E(\P)\), there is an element of `choice' associated to the Lyapunov-like functions \(V_{i}\), \(i\in\P\) and hence the corresponding scalars \(\lambda_{i}\), \(i\in\P\) and \(\mu_{ij}\), \((i,j)\in E(\P)\). A natural question is, therefore, the following: given a family of systems and a set of admissible switches, can we design Lyapunov-like functions \(V_{i}\), \(i\in\P\) (and hence, the corresponding scalars \(\lambda_{i}\), \(i\in\P\) and \(\mu_{ij}\), \((i,j)\in E(\P)\)) such that \(\overline{G}(V,E)\) admits a negative weight cycle? To the best of our knowledge, this question is numerically difficult, and only a partial solution can be obtained as a special case of \cite[Algorithm 2]{KunQue_TCNS_2019}. This partial solution searches for suitable scalars over a finite number of choices from the set of quadratic Lyapunov-like functions, see also \cite[Remarks 9 and 12]{KunQue_TCNS_2019} for discussions.

        In the absence of a complete solution to the above question, the existing literature considers the Lyapunov-like functions \(V_{i}\), \(i\in\P\) and the corresponding scalars \(\lambda_{i}\), \(i\in\P\) and \(\mu_{ij}\), \((i,j)\in E(\P)\) to be ``given'', and searches for negative weight cycles on \(\overline{G}(V,E)\). However, non-existence of such a cycle with the given sets of vertex and edge weights, does not conclude that \eqref{e:family} does not admit such a cycle at all, and hence adds to the conservatism of the class of results under consideration. Moreover, for large-scale switched systems, storing the vertex and edge weights prior to the application of a negative weight cycle detection algorithm, has a large memory requirement. Recently in \cite{BalaKunCha_MCRF_2019} the authors proposed a randomized algorithm that detects a cycle on \(\overline{G}(V,E)\) without prior knowledge of vertex and edge weights of the graph, and identified sufficient conditions on the connectivity and vertex and edge weights of \(\overline{G}(V,E)\) under which such a cycle is stabilizing (in the sense that it is a  negative weight cycle). The conditions on favourable vertex and edge weights are provided in terms of their statistical properties. However, given a family of systems \eqref{e:family}, the problem of designing Lyapunov-like functions \(V_{i}\), \(i\in\P\) such that the corresponding scalars \(\lambda_{i}\), \(i\in\P\) and \(\mu_{ij}\), \((i,j)\in E(\P)\) satisfy the proposed conditions is not addressed.

        In the current work we employ properties of the subsystem matrices and avoid the use of Lyapunov-like functions. In contrast to \(\overline{G}(V,E)\), we do not associate weights to the vertices and edges of \(G(V,E)\), and our detection mechanism for a set cycles, \(C_{p}\), solely depends on the connectivity of \(G(V,E)\). However, {the stability conditions} presented in this paper are restricted to the setting of switched linear systems unlike the case of Lyapunov-like functions based {construction} of stabilizing cycles that extends to the nonlinear setting, as demonstrated in \cite{KunCha_NAHS_2017}.
\section{Concluding remarks}
\label{s:concln}
    In this paper we proposed a new {method to construct} stabilizing cycles for switched linear systems. The proposed result involves properties of commutators of the subsystem matrices and mild assumption on the admissible switches between the subsystems. {Algorithmic construction} of stabilizing cycles for switched nonlinear systems without involving the design of Lyapunov-like functions remains an open question.
\section{Proof of our result}
\label{s:all_proofs}
    {\it Proof of Theorem \ref{t:mainres}}: Let \eqref{e:family} admit a \(p\in\P_{S}\) for which conditions \eqref{e:maincondn2} and \eqref{e:maincondn3} hold. Let \(\sigma\) be a switching signal whose corresponding infinite walk is constructed by concatenating elements from the set \(C_{p}\). Let \(\M\) be the corresponding word (matrix product) defined as
    \(
        \M = \cdots A_{\sigma(2)}A_{\sigma(1)}A_{\sigma(0)}.
    \)
    The condition \eqref{e:ges1} for GES of \eqref{e:swsys} under \(\sigma\) can be written equivalently as \cite[\S 2]{Agrachev2012}: there exist positive numbers \(c\) and \(\gamma\) such that
    \begin{align}
    \label{e:pf1_step1}
    	\norm{\M} \leq c e^{-\gamma\abs{\M}}\:\:\text{for all}\:\abs{\M}.
    \end{align}

    It, therefore, suffices to show that condition \eqref{e:pf1_step1} holds for the above \(\sigma\). We will employ mathematical induction on \(\abs{\M}\) for this purpose.

    {\it A. Induction basis}: Pick \(c\) large enough so that \eqref{e:pf1_step1} holds for \(\M\) satisfying \(\abs{\M}\leq Nm\).

    {\it B. Induction hypothesis}: Let \(\abs{\M}\geq Nm + 1\) and assume that \eqref{e:pf1_step1} is proved for all products of length less than \(\abs{\M}\).

    {\it C. Induction step}: Let \(\M = LR\), where \(\abs{L} = Nm\). We observe that \(L\) contains at least \(m\)-many \(A_{p}\). Indeed, each cycle in \(C_{p}\) contains vertex \(p\), and the length of any cycle on \(G(V,E)\) is at most \(N\).

    We rewrite \(L\) as
    \(
    	L = A_{p}^{m}L_{1} + L_{2},
    \)
    where \(\abs{L_{1}} = (N-1)m\), and \(L_{2}\) contains at most \((N-1)\frac{m(m+1)}{2}\) terms of length \(Nm-1\) with \(Nm-2\) \(A_{i}\), \(i\in\P\) and \(1\) \(E_{p,i}\), \(i\in\P\setminus\{p\}\).

    Now, from the sub-multiplicativity and sub-additivity properties of the induced norm, we have
    \begin{align}
    \label{e:pf1_step2}
    	&\hspace*{-0.15cm}\norm{\M} = \norm{LR}  \leq \norm{A_{p}^{m}}\norm{L_{1}R} + \norm{L_{2}}\norm{R}\nonumber\\
	&\hspace*{-0.15cm}\leq \rho ce^{-\gamma(\abs{\M}-m)} + (N-1)\frac{m(m+1)}{2}M^{Nm-2}\varepsilon_{p} ce^{-\gamma(\abs{\M}-Nm)}\nonumber\\
	&\hspace*{-0.15cm}=ce^{-\gamma\abs{\M}}\bigl(\rho e^{\lambda m} + (N-1)({m(m+1)}/{2})M^{Nm-2}\varepsilon_{p}
	e^{\gamma Nm}\bigr).
    \end{align}
    In the above inequality the upper bounds on \(\norm{L_{1}R}\) and \(\norm{R}\) are obtained from the relations \(\abs{\M} = \abs{A_{p}^{m}}+\abs{L_{1}R}\) and \(\abs{\M} = \abs{L}+\abs{R}\), respectively. Applying \eqref{e:maincondn3} to \eqref{e:pf1_step2} leads to \eqref{e:pf1_step1}. Consequently, \eqref{e:swsys} is GES under the \(\sigma\) in consideration.

    This completes our proof of Theorem \ref{t:mainres}.

    \begin{remark}
    \label{rem:proof_tech}
    	Our proof of Theorem \ref{t:mainres} relies on combinatorial arguments applied to the matrix product \(\M\) split into sums \(A_{p}^{m}L_{1}R + L_{2}R\). Combinatorial analysis technique for stability of switched system was first introduced in \cite{Agrachev2012} for stability under arbitrary switching, and was recently extended to stability under pre-specified dwell times in \cite{abc,def}. The following properties of stabilizing cycles are used in our analysis: (i) each of them contains a vertex \(p\) for which conditions \eqref{e:maincondn2}-\eqref{e:maincondn3} hold, and (ii) each of them is of length at most \(N\). The choice of matrix commutators involves the subsystem matrix \(A_{p}\) and the subsystem matrices \(A_{i}\), \(i\in\P\setminus\{p\}\), and they are utilized to rearrange any possible product corresponding to an infinite walk constructed by concatenating elements from \(C_{p}\). Notice that the upper bound on the number of terms in \(L_{2}\) and their structure are obtained by exchanging \(A_{p}\) with \(A_{i}\), \(i\in\P\setminus\{p\}\) towards achieving the form \(L = A_{p}^{m}L_{1}+L_{2}\). Consider, for example, \(G(V,E)\) with \(V = \{1,2,3,4\}\), \(\P_{S} = \{1,2\}\), \(\P_{U} = \{3,4\}\) and \(E(\P) = \{(1,2),(1,3),(1,4),(2,1),(2,3),(3,4),(4,1),(4,2),(4,3)\}\). Let \(m=2\), and conditions \eqref{e:maincondn2}-\eqref{e:maincondn3} hold for \(p=1\). Suppose that \(L = A_{2}A_{4}A_{3}A_{1}A_{4}A_{3}A_{2}A_{1}\).
    It can be rewritten as
    \begin{align*}
        &A_{2}A_{4}\underline{A_{3}A_{1}}A_{4}A_{3}A_{2}A_{1}\\
        =& A_{2}\underline{A_{4}A_{1}}A_{3}A_{4}A_{3}A_{2}A_{1}-A_{2}A_{4}E_{3,1}A_{4}A_{3}A_{2}A_{1}\\
        =& \underline{A_{2}A_{1}}A_{4}A_{3}A_{4}A_{3}A_{2}A_{1}-A_{2}E_{4,1}A_{3}A_{4}A_{3}A_{2}A_{1} 
        - A_{2}A_{4}E_{3,1}A_{4}A_{3}A_{2}A_{1}\\
        =& A_{1}A_{2}A_{4}A_{3}A_{4}A_{3}\underline{A_{2}A_{1}}-E_{2,1}A_{4}A_{3}A_{4}A_{3}A_{2}A_{1}
        -A_{2}E_{4,1}A_{3}A_{4}A_{3}A_{2}A_{1}\\
        &- A_{2}A_{4}E_{3,1}A_{4}A_{3}A_{2}A_{1}\\
        =&A_{1}A_{2}A_{4}A_{3}A_{4}\underline{A_{3}A_{1}}A_{2}-A_{1}A_{2}A_{4}A_{3}A_{4}A_{3}E_{2,1}
        -E_{2,1}A_{4}A_{3}A_{4}A_{3}A_{2}A_{1}\\
        &-A_{2}E_{4,1}A_{3}A_{4}A_{3}A_{2}A_{1}- A_{2}A_{4}E_{3,1}A_{4}A_{3}A_{2}A_{1}\\
        =&A_{1}A_{2}A_{4}A_{3}\underline{A_{4}A_{1}}A_{3}A_{2}-A_{1}A_{2}A_{4}A_{3}A_{4}E_{3,1}A_{2}
        -A_{1}A_{2}A_{4}A_{3}A_{4}A_{3}E_{2,1}\\
        &-E_{2,1}A_{4}A_{3}A_{4}A_{3}A_{2}A_{1}
        -A_{2}E_{4,1}A_{3}A_{4}A_{3}A_{2}A_{1} - A_{2}A_{4}E_{3,1}A_{4}A_{3}A_{2}A_{1}\\
        =&A_{1}A_{2}A_{4}\underline{A_{3}A_{1}}A_{4}A_{3}A_{2}-A_{1}A_{2}A_{4}A_{3}E_{4,1}A_{3}A_{2}
        -A_{1}A_{2}A_{4}A_{3}A_{4}E_{3,1}A_{2}\\
        &-A_{1}A_{2}A_{4}A_{3}A_{4}A_{3}E_{2,1}-E_{2,1}A_{4}A_{3}A_{4}A_{3}A_{2}A_{1}-A_{2}E_{4,1}A_{3}A_{4}A_{3}A_{2}A_{1}\\
        &- A_{2}A_{4}E_{3,1}A_{4}A_{3}A_{2}A_{1}\\
        =&A_{1}A_{2}\underline{A_{4}A_{1}}A_{3}A_{4}A_{3}A_{2}-A_{1}A_{2}A_{4}E_{3,1}A_{4}A_{3}A_{2}
        -A_{1}A_{2}A_{4}A_{3}E_{4,1}A_{3}A_{2}\\
        &-A_{1}A_{2}A_{4}A_{3}A_{4}E_{3,1}A_{2}-A_{1}A_{2}A_{4}A_{3}A_{4}A_{3}E_{2,1}-E_{2,1}A_{4}A_{3}A_{4}A_{3}A_{2}A_{1}\\
        &-A_{2}E_{4,1}A_{3}A_{4}A_{3}A_{2}A_{1} - A_{2}A_{4}E_{3,1}A_{4}A_{3}A_{2}A_{1}\\
        =&A_{1}\underline{A_{2}A_{1}}A_{4}A_{3}A_{4}A_{3}A_{2}-A_{1}A_{2}E_{4,1}A_{3}A_{4}A_{3}A_{2}
        -A_{1}A_{2}A_{4}E_{3,1}A_{4}A_{3}A_{2}\\
        &-A_{1}A_{2}A_{4}A_{3}E_{4,1}A_{3}A_{2}
        -A_{1}A_{2}A_{4}A_{3}A_{4}E_{3,1}A_{2}-A_{1}A_{2}A_{4}A_{3}A_{4}A_{3}E_{2,1}\\
        &-E_{2,1}A_{4}A_{3}A_{4}A_{3}A_{2}A_{1}-A_{2}E_{4,1}A_{3}A_{4}A_{3}A_{2}A_{1}
        - A_{2}A_{4}E_{3,1}A_{4}A_{3}A_{2}A_{1}\\
        =&{A_{1}^{2}}A_{2}A_{4}A_{3}A_{4}A_{3}A_{2}-A_{1}E_{2,1}A_{4}A_{3}A_{4}A_{3}A_{2}
        -A_{1}A_{2}E_{4,1}A_{3}A_{4}A_{3}A_{2}\\
        &-A_{1}A_{2}A_{4}E_{3,1}A_{4}A_{3}A_{2}-A_{1}A_{2}A_{4}A_{3}E_{4,1}A_{3}A_{2}-A_{1}A_{2}A_{4}A_{3}A_{4}E_{3,1}A_{2}\\
        &-A_{1}A_{2}A_{4}A_{3}A_{4}A_{3}E_{2,1}-E_{2,1}A_{4}A_{3}A_{4}A_{3}A_{2}A_{1}-A_{2}E_{4,1}A_{3}A_{4}A_{3}A_{2}A_{1}\\
        &- A_{2}A_{4}E_{3,1}A_{4}A_{3}A_{2}A_{1}.
    \end{align*}
    In the worst case the \(k\)-th instance of \(A_{p}\) in \(L\) (reading from the right) is to be exchanged with \(k(N-1)\) terms, \(k = 1,2,\ldots,m\), to obtain the desired structure of \(L\).
    \end{remark}


\end{document}